\documentclass[12pt]{article}
\usepackage{amsmath}
\usepackage{amssymb}
\usepackage{amsfonts}

\DeclareMathOperator{\sech}{sech}
\DeclareMathOperator{\arctanh}{arctanh}

\title{ 
Comment on `Exact solution of the position-dependent effective mass and angular frequency Schr\"odinger 
equation: harmonic oscillator model with quantized confinement parameter'}
\author{C Quesne\\ 
{\small Physique Nucl\'eaire Th\'eorique et Physique Math\'ematique,  Universit\'e Libre de Bruxelles,} \\ 
{\small Campus de la Plaine CP229, Boulevard~du Triomphe, B-1050 Brussels, Belgium}\\
{\small E-mail: Christiane.Quesne@ulb.be}}
\date{ }
\begin{document}
\baselineskip=20pt plus 1pt minus 1pt
\maketitle

\begin{abstract}
In a recent paper by Jafarov, Nagiyev, Oste and Van der Jeugt (2020 {\sl J.\ Phys.\ A} {\bf 53} 485301), a confined model of the non-relativistic quantum harmonic oscillator, where the effective mass and the angular frequency are dependent on the position, was constructed and it was shown that the confinement parameter gets quantized. By using a point canonical transformation starting from the constant-mass Schr\"odinger equation for the Rosen-Morse II potential, it is shown here that similar results can be easily obtained without quantizing the confinement parameter. In addition, an extension to a confined shifted harmonic oscillator directly follows from the same point canonical transformation.
\end{abstract}


%
%
\newpage

In an interesting paper \cite{jafarov}, Jafarov {\sl et al} presented an exact solution of a confined model of the non-relativistic quantum harmonic oscillator, where the effective mass and the angular frequency are dependent on the position. In their model, the kinetic energy operator has the BenDaniel-Duke form \cite{bendaniel} and the position-dependency of the mass and the angular frequency is such that the homogeneous nature of the harmonic oscillator force constant $k = M(x) \omega^2(x) = m_0 \omega_0^2$ and hence the regular harmonic oscillator potential are preserved, thus leading to the Schr\"odinger equation \footnote{For simplicity's sake, we use here units wherein $\hbar = 2m_0 = 1$.}
\begin{equation}
  \left(- \frac{d}{dx} \frac{1}{M(x)} \frac{d}{dx} + V_{\rm eff}(x)\right) \psi_n(x) = E_n \psi_n(x)
  \label{eq:PDM-eq}
\end{equation}
with
\begin{equation}
  M(x) = \left(1 - \frac{x^2}{a^2}\right)^{-2}, \qquad V_{\rm eff}(x) = \frac{1}{4} \omega_0^2 x^2, \qquad
  -a < x < a.  \label{eq:PDM-V}
\end{equation}
The equation is solved by reducing it to the general Legendre equation. As a consequence, a quantization of the confinement parameter $a$ is observed,
\begin{equation}
  a = a_l = \sqrt{\frac{2}{\omega_0}}\, [l(l+1)-2]^{1/4}, \qquad l = 2, 3, \dots,
\end{equation}
the bound-state wavefunctions being expressed in terms of associated Legendre polynomials or, equivalently, Gegenbauer polynomials,
\begin{align}
  &\psi_n(x) = N_n \left(1-\frac{x^2}{a^2}\right)^{(l-n-1)/2} C_n^{\left(l-n+\frac{1}{2}\right)} 
  \left(\frac{x}{a}\right), \\
  &N_n = \frac{(2l-2n)!}{2^{l-n}(l-n)!} \left(\frac{(l-n) n!}{a (2l-n)!}\right)^{1/2},
\end{align}
with the corresponding $E_n$ given by
\begin{equation}
  E_n = \omega_0 \sqrt{1+ \left(\frac{3}{\omega_0 a^2}\right)^2}\, \left(n+\frac{1}{2}\right) - \frac{1}{a^2}
  \left(n+\frac{1}{2}\right)^2 - \frac{5}{4a^2}, \qquad n=0, 1, \ldots, l-2.
\end{equation}
\par
%
%
The purpose of the present comment is to show that an alternative derivation of the solution consists in applying a point canonical transformation  (PCT) to the well-known constant-mass Schr\"odinger equation for the Rosen-Morse II potential \cite{rosen, levai89, cooper, levai09}. Such an approach has the advantage of providing us with a direct generalization of the results.\par
%
%
In the PCT approach to position-dependent mass (PDM) Schr\"odinger equations \cite{bagchi, cq}, one starts from a constant-mass Schr\"odinger equation
\begin{equation}
  \left(- \frac{d^2}{du^2} + U(u)\right) \phi_n(u) = \epsilon_n \phi_n(u),  \label{eq:CM-eq}
\end{equation}
for some potential $U(u)$, defined on a finite or infinite interval, and one transforms it into an equation such as (\ref{eq:PDM-eq}) by making a change of variable
\begin{equation}
  u(x) = \bar{a} v(x) + \bar{b}, \qquad v(x) = \int^x \sqrt{M(x')}\, dx',  \label{eq:u-v}
\end{equation}
and a change of function
\begin{equation}
  \phi_n(u(x)) \propto [M(x)]^{-1/4} \psi_n(x).  \label{eq:phi-psi}
\end{equation}
Here, $\bar{a}$ and $\bar{b}$ are assumed to be two real parameters. The potential $V_{\rm eff}(x)$, defined on a possibly different interval, and the energy eigenvalues $E_n$ of the PDM Schr\"odinger equation are given in terms of the potential and the energy eigenvalues of the constant-mass one by
\begin{equation}
  V_{\rm eff}(x) = \bar{a}^2 U(u(x)) + \frac{M^{\prime\prime}}{4M^2} - \frac{7 M^{\prime2}}{16M^3} +
  \bar{c}. \label{eq:V-U}
\end{equation}
and
\begin{equation}
  E_n = \bar{a}^2 \epsilon_n + \bar{c},  \label{eq:E-epsilon}
\end{equation}
where a prime denotes derivation with respect to $x$ and $\bar{c}$ is some additional real constant.\par
%
%
In equation (\ref{eq:CM-eq}), let us assume that $U(u)$ is the Rosen-Morse II potential
\begin{equation}
  U(u) = - A(A+1) \sech^2 u + 2B \tanh u, \qquad -\infty<u<\infty, \qquad B<A^2.  \label{eq:RM-pot}
\end{equation}
Then the corresponding $\epsilon_n$ and $\phi_n(u)$ are given by
\begin{equation}
  \epsilon_n = - (A-n)^2 - \frac{B^2}{(A-n)^2}, \qquad n=0, 1, \ldots, n_{\rm max}, \qquad n_{\rm max} <
  A - \sqrt{|B|},
\end{equation}
and
\begin{align}
  \phi_n(u) &= {\cal N}_n (1-\tanh u)^{\left(A-n+\frac{B}{A-n}\right)/2} (1+\tanh u)^{\left(A-n-\frac{B}{A-n}
      \right)/2} \nonumber \\
  & \quad \times P_n^{\left(A-n + \frac{B}{A-n}, A-n - \frac{B}{A-n}\right)}(\tanh u),
\end{align} 
with $P_n^{(\alpha,\beta)}(z)$ denoting a Jacobi polynomial and ${\cal N}_n$ a normalization coefficient given by \cite{levai09}
\begin{equation}
  {\cal N}_n = 2^{n-A} \left(\frac{n! \Gamma(2A-n+1) [(A-n)^2 - B^2/(A-n)^2]}{(A-n) \Gamma(A+1+B/(A-n))
  \Gamma(A+1-B/(A-n))}\right)^{1/2}
\end{equation}
and ensuring that
\begin{equation}
  \int_{-\infty}^{+\infty} |\phi_n(u)|^2 = 1.
\end{equation}
%
%
To start with, let us consider the special case where $B=0$ in equation (\ref{eq:RM-pot}). Then, for $\epsilon_n = - (A-n)^2$, $n=0$, 1, \ldots, $n_{\rm max}$, $n_{\rm max} < A$, the wavefunctions $\phi_n(u)$ can be expressed in terms of Gegenbauer polynomials as
\begin{equation}
  \phi_n(u) = \frac{\Gamma(2A-2n+1)}{2^{A-n} \Gamma(A-n+1)} \left(\frac{(A-n)n!}{\Gamma(2A-n+1)}
  \right)^{1/2} (\sech u)^{A-n}\, C_n^{\left(A-n+\frac{1}{2}\right)}(u). \label{eq:phi}
\end{equation}
by taking into account equation (22.5.20) of Ref.~\cite{abramowitz}, expressing Jacobi polynomials with equal parameters in terms of Gegenbauer ones.\par
%
%
{}For the PDM $M(x)$ defined in equation (\ref{eq:PDM-V}), we get
\begin{equation}
  v(x) = a \arctanh \frac{x}{a}
\end{equation}
and
\begin{equation}
  \frac{M^{\prime\prime}}{4M^2} - \frac{7M^{\prime2}}{16M^3} = - \frac{2x^2}{a^4} + \frac{1}{a^2}.
\end{equation}
With the choice
\begin{equation}
  \bar{a} = \frac{1}{a}, \qquad \bar{b}=0
\end{equation}
for the constants appearing in equation (\ref{eq:u-v}), we obtain for the change of variable
\begin{equation}
  u(x) = \arctanh \frac{x}{a}.
\end{equation}
On selecting 
\begin{equation}
  \bar{c} = \frac{1}{4} \omega_0^2 a^2 + \frac{1}{a^2},  \label{eq:c}
\end{equation}
the transformed potential (\ref{eq:V-U}) becomes
\begin{equation}
  V_{\rm eff}(x) = \frac{1}{a^4}[A(A+1)-2] x^2 - \frac{1}{a^2}[A(A+1)-2] + \frac{1}{4}\omega_0^2 a^2
\end{equation}
and reduces to that given in equation (\ref{eq:PDM-V}) by choosing
\begin{equation}
  a = \sqrt{\frac{2}{\omega_0}}\, [A(A+1)-2]^{1/4}  \label{eq:a}
\end{equation}
provided $A(A+1)>2$ or, in other words, $A>1$. From equations (\ref{eq:phi-psi}) and (\ref{eq:phi}), the corresponding bound-state wavefunctions, normalized on $(-a, +a)$, can be written as
\begin{align}
  \psi_n(x) &= \frac{\Gamma(2A-2n+1)}{2^{A-n} \Gamma(A-n+1)} \left(\frac{(A-n) n!}{a \Gamma(2A-n+1)}
       \right)^{1/2} \left(1 - \frac{x^2}{a^2}\right)^{(A-n-1)/2} \nonumber \\
  & \quad \times C_n^{\left(A-n+\frac{1}{2}\right)}\left(\frac{x}{a}\right),  \label{eq:psi}
\end{align}
with $E_n$ obtained from (\ref{eq:E-epsilon}) as
\begin{align}
  E_n &= - \frac{1}{a^2} (A-n)^2 + \frac{1}{4} \omega_0^2 a^2  + \frac{1}{a^2} \nonumber \\
  &= \omega_0 \sqrt{1+ \left(\frac{3}{\omega_0 a^2}\right)^2} \left(n+\frac{1}{2}\right) - \frac{1}{a^2}
      \left(n+ \frac{1}{2}\right)^2 - \frac{5}{4a^2}, \nonumber \\
  & \quad n=0, 1, \ldots, n_{\rm max}, \qquad n_{\rm max} < A-1.
\end{align}
Note that the condition $n=0, 1, \ldots, n_{\rm max}$, $n_{\rm max} < A$, which was enough to make the functions $\phi_n(u)$ of equation (\ref{eq:phi}) normalizable on $(-\infty, +\infty)$, has been restricted to $n=0, 1, \ldots, n_{\rm max}$, $n_{\rm max} < A-1$, which is imposed  by the normalizability condition of $\psi_n(x)$ on $(-a,+a)$.\par
%
%
We conclude that the results of Ref.~\cite{jafarov} can be derived by a PCT applied to the constant-mass Schr\"odinger equation for the Rosen-Morse II potential with $B=0$ and any $A= l = 2, 3, \ldots$. However, no quantization of $A$, and hence of $a$, is necessary to get valid bound-state wavefunctions: for any positive $A$ greater than one, we have shown that equation (\ref{eq:PDM-eq}) with $M(x)$ and $V_{\rm eff}(x)$ given in equation (\ref{eq:PDM-V}) has a finite number of bound-state wavefunctions (\ref{eq:psi}).\par
%
%
Let us now consider the general Rosen-Morse II potential with $B\ne 0$. On proceeding as above, except for changing $\bar{c}$ of equation (\ref{eq:c}) into
\begin{equation}
  \bar{c} = \frac{1}{4} \omega_0^2 a^2 + \frac{1}{a^2} + b^2,
\end{equation}
after setting $B = - \frac{1}{2} \omega_0 a^3 b$, we get
\begin{equation}
  V_{\rm eff}(x) = \frac{1}{4} \omega_0^2 \left(x - \frac{2b}{\omega_0}\right)^2, \qquad -a<x<a,
\end{equation}
with $a$ still given by equation (\ref{eq:a}). The confined harmonic oscillator of equation (\ref{eq:PDM-V}) is therefore replaced by a confined shifted harmonic oscillator. The latter has a finite number of bound-state wavefunctions
\begin{align}
  \psi_n(x) &= N_n \left(1-\frac{x}{a}\right)^{\left(A-n-1- \frac{\omega_0 a^3 b}{2(A-n)}\right)/2}
      \left(1+\frac{x}{a}\right)^{\left(A-n-1+ \frac{\omega_0 a^3 b}{2(A-n)}\right)/2} \nonumber \\
  &\quad \times P_n^{\left(A-n - \frac{\omega_0 a^3 b}{2(A-n)}, A-n + \frac{\omega_0 a^3 b}{2(A-n)}
      \right)} \left(\frac{x}{a}\right), \\
  N_n &= 2^{n-A} \Bigl(n! \Gamma(2A-n+1) \{(A-n)^2 - \omega_0^2 a^6 b^2/[4(A-n)^2]\}\Bigr)^{1/2}
      \nonumber \\
  &\quad \times \Bigl(a (A-n) \Gamma(A+1 - \omega_0 a^3 b/[2(A-n)]) 
      \Gamma(A+1 + \omega_0 a^3 b/[2(A-n)])\Bigr)^{-1/2},
\end{align}
with $n=0$, 1, \ldots, $n_{\rm max}$, $n_{\rm max} < A - \frac{1}{2} \left(1 + \sqrt{1 + 2\omega_0 a^3 |b|}\right)$, provided $b$ satisfies the condition $\sqrt{1 + 2\omega_0 a^3 |b|} < 2A-1$, which on taking (\ref{eq:a}) into account, amounts to
\begin{equation}
  |b| < \sqrt{\frac{\omega_0}{2}} \frac{A(A-1)}{[A(A+1)-2]^{3/4}}.
\end{equation}
The corresponding bound-state energies, as obtained from equation (\ref{eq:E-epsilon}), are given by
\begin{equation}
  E_n = \omega_0 \sqrt{1 + \left(\frac{3}{\omega_0 a^2}\right)^2} \left(n+\frac{1}{2}\right) - \frac{1}{a^2}
  \left(n+\frac{1}{2}\right)^2 - \frac{5}{4a^2} + b^2 \frac{g(n)}{f(n)},
\end{equation}
where
\begin{align}
  f(n) &=\left(n+\frac{1}{2}\right)^2 - \omega_0 a^2 \sqrt{1+ \left(\frac{3}{\omega_0 a^2}\right)^2}
      \left(n+\frac{1}{2}\right) + \frac{1}{4} \omega_0^2 a^4 + \frac{9}{4}, \\
  g(n) &= f(n) - \frac{1}{4} \omega_0^2 a^4. 
\end{align}
\par
%
%
In conclusion, we have shown that the confined harmonic oscillator with position-dependent  mass and angular frequency of Ref.~\cite{jafarov} can be extended to the case where the confinement parameter is not quantized, its spectrum remaining finite, non-equidistant, and dependent on the confinement parameter. A further extension to a confined shifted harmonic oscillator with similar characteristics has also been constructed.\par
%
%
This work was supported by the Fonds de la Recherche Scientifique - FNRS under Grant Number 4.45.10.08.\par
%
%

 \end{document}